\documentclass[useAMS,usenatbib,usegraphicx]{mn2e}
\usepackage{times}

\newif\ifAMStwofonts
\AMStwofontstrue

\newcommand{\simlt}{\lower.5ex\hbox{$\; \buildrel < \over \sim \;$}}
\newcommand{\simgt}{\lower.5ex\hbox{$\; \buildrel > \over \sim \;$}}
\newcommand{\be}{\begin{equation}}
\newcommand{\ba}{\begin{eqnarray}}
\newcommand{\ee}{\end{equation}}
\newcommand{\ea}{\end{eqnarray}}

\title[PCA and the SFH of elliptical galaxies]
{A Principal Component Analysis approach to the Star Formation History
of elliptical galaxies in Compact Groups}
\author[I. Ferreras et al.]
{Ignacio Ferreras$^{1,2}$\thanks{E-mail: ferreras@star.ucl.ac.uk},
Anna Pasquali$^3$, Reinaldo R. de~Carvalho$^4$, 
\newauthor
Ignacio G. de~la~Rosa$^5$ and Ofer Lahav$^1$\\
$^1$ Department of Physics and Astronomy, University College London,
  Gower St. London WC1E 6BT\\
$^2$ Department of Physics, King's College London, Strand, London WC2R 6LS\\
$^3$ Max-Planck-Institut f\"ur Astronomie, Koenigstuhl 17, D-69117 Heidelberg, Germany\\
$^4$ INPE/MCT, Avenida dos Astronautas 1758, S\~ao Jos\'e dos Campos, SP 12227--010, Brazil\\
$^5$ Instituto de Astrof\'\i sica de Canarias, Calle V\'\i a L\'actea,
  E-38200 La Laguna, Tenerife, Spain
}

\begin{document}
\date{Accepted version MNRAS 2006 May 3.}
\pagerange{\pageref{firstpage}--\pageref{lastpage}} \pubyear{2006}
\maketitle
\label{firstpage}

\begin{abstract}
Environmental differences in the stellar populations of early-type
galaxies are explored using principal component analysis (PCA),
focusing on differences between elliptical galaxies in Hickson Compact
Groups (HCGs) and in the field. The method is model-independent and
purely relies on variations between the observed spectra. The
projections (PC1,PC2) of the observed spectra on the first and second
principal components reveal a difference with respect to environment,
with a wider range in PC1 and PC2 in the group sample. We define a
spectral parameter ($\zeta\equiv 0.36$PC1$-$PC2) which simplifies this
result to a single number: field galaxies have a very similar
value of $\zeta$, whereas HCG galaxies span a wide range in this
parameter.  The segregation is found regardless of the way the input
SEDs are presented to PCA (i.e. changing the spectral range; using
uncalibrated data; subtracting the continuum or masking the SED to
include only the Lick spectral regions). Simple models are applied to
give physical meaning to the PCs. We obtain a strong correlation
between the values of $\zeta$ and the mass fraction in younger stars,
so that some group galaxies present a higher fraction of them,
implying a more complex star formation history in groups. 
Regarding ``dynamically-related'' observables such as $a_4$ or
velocity dispersion, we find a correlation with PC3, but not with
either PC1 or PC2. PCA is more sensitive than other methods based on
a direct analysis of observables such as the structure of the surface
brightness profile or the equivalent width of absorption lines. The
latter do not reveal any significant variation between field and
compact group galaxies. Our results imply that the 
presence of young stars only amounts to a fraction of a percent
in its contribution to the total variance, reflecting the power of PCA
as a tool to extract small variations in the spectra from unresolved
stellar populations.
\end{abstract}


\section{Introduction}

The connection between mass assembly and star formation provides the
best approach to building a comprehensive picture of galaxy
formation. In this respect, massive elliptical galaxies are ideal
probes since their light is mostly dominated by old stellar
populations, whereas their mass and morphology require late
assembly. Both old ages and late assembly can be reconciled within the
standard concordance cosmology. However, we currently lack a
method to give an accurate estimate of the star formation history
(SFH) from photo-spectroscopic data.

There are various factors that contribute to ``decouple'' the
evolution of the total and stellar mass content of a galaxy. Most
notably, the effect of feedback on star formation can make these two
components follow divergent paths: for instance, a highly
efficient mechanism of star formation at early stages can lead to
``dry mergers'' at later times (Bell et al. 2004).  Indeed, the
so-called downsizing trend (see e.g. Cowie et al. 1996; Treu et
al. 2005) -- in which lower mass galaxies contain most of the global
star formation at later times -- supports this effect.

The environment is another factor which can play an important
role in shaping the SFHs of galaxies. Clusters of galaxies trace high
density regions that collapsed earlier and present galaxies with a
``hostile'' environment in which gas can be removed by, for instance,
ram pressure, galaxy harassment, and tidal interaction (e.g. Haynes
1989).  Furthermore, the high virial velocities in clusters prevent
mergers from taking place. This effect is dependent on the cosmic
epoch and might be a non-trivial one to accurately estimate. The
association between compact groups (CGs) and clusters has been
investigated by Rood \& Struble (1994) who find that 75\% of Hickson's
groups seem to be associated to structures such as loose groups and
clusters, indicating that CGs are actually part of the same observed
hierarchy from isolated galaxies to superclusters.  More recently, de
Carvalho et al. (2005) examining a sample of CGs at a slightly larger
redshift than Hickson's find that there is an excess of CGs within one
Abell radius of the nearest cluster, over a random distribution
(32\%). Moreover, they find a marginal excess of CGs related to rich
clusters relative to poor clusters, which might be important for
considering properly the environment where CGs live in. Therefore, it
is of paramount importance to establish the connection between CGs and
the large scale structure. In this context, it is important to remember
that groups are the dominant type of structure found in the Universe
(Nolthenius \& White 1987) making them especially important from the
cosmological viewpoint.  

The effects of the environment on the galaxian properties have been
studied over more than two decades (Guzman et al 1991; de Carvalho \&
Djorgovski 1992) and the issue is still debatable. Recent work
regarding differences between early-type galaxies with respect to
environment have found controversial results.  While some indicate
that environment does play some role (de Carvalho \& Djorgovski 1992;
Ziegler et al. 2005), others find no difference at all (e.g. de la
Rosa et al 2001a; Bernardi et al. 1998). Also, studies of the stellar
population of early-type systems indicate that they are older and and
more metal poor in compact groups and clusters than in the field (Rose
et al. 1994; Proctor et al. 2005; de la Rosa et al. 2001b; Mendes de
Oliveira et al. 2005), suggesting that Es in dense systems experience
a truncated period of star formation, which ultimately affects their
chemical enrichment history, as opposed to their counterparts in the
low density regime. 

This paper presents an alternative approach to the analysis of
unresolved old stellar populations, focused towards a comparison of
the star formation history between CGs and the field. The current
epoch of large surveys provides a fertile ground for statistical
techniques such as PCA (Madgwick et al. 2002), ICA (Lu et al. 2006),
IB (Slonim et al. 2001) or ANNs (Naim et al. 1995). We show here the
capabilities of PCA to disentangle the star formation history of
early-type galaxies. The structure of the paper is as follows: \S2
presents the sample, \S3 explains the technique of principal component
analysis applied to spectra. \S4 gives the results for our sample,
which are put in context with physical models in \S5, followed by a
discussion in \S6.

\section{The sample}
Our sample comprises 30 elliptical galaxies split into 18 located in
the cores of Hickson Compact Groups (HCGs; Hickson 1982), and a
control set of 12 early-type galaxies located either in the field, in
the outskirts of clusters or in very loose groups.  The sample selection
criteria and spectroscopic
data is presented in de~la~Rosa, de~Carvalho \& Zepf (2001a). The main
details of the sample are listed in table~\ref{tab:sample}.  All
galaxies were observed with the same instrument and configuration:
long-slit spectroscopy over the 3500--7000\AA\ range, with a FWHM
resolution of 4.25\AA\ . We removed from the original sample four
galaxies (HCG 28b; 37e; 46a; 59b) since they have rather low
signal-to-noise ratios. The remaining galaxies in the sample have
S/N$\simgt 35$\AA$^{-1}$. The extraction aperture was chosen to
encompass all available flux (de~la~Rosa et al. 2001a). Flux
calibration was performed by a comparison with standard stars observed
during the same run and with the same configuration.  Three
standard stars were used: Hiltner 600 (B3), HZ44 (B2) and Feige 34 (DA).
Comparing the flux-calibrated spectra using independently these stars, 
we get discrepancies in flux up to 7\%.

The SEDs were dereddened using the Fitzpatrick (1999) $R_v=3.1$
Galactic extinction curve, taking the reddening values from the maps of
Schlegel, Finkbeiner \& Davis (1998) as shown in
table~\ref{tab:sample}.  Notice that the reddening
values between group and field sample span a similar range. 

\begin{table*}
\begin{minipage}{15cm}
\caption{Group and Field sample}
\label{tab:sample}
\begin{tabular}{lcccccrrrr}
\hline\hline
Galaxy & Type & $M_B$ & $\log\sigma_0$ & E(B$-$V) & 
PC1$^a$ & PC2$^a$ & PC3$^a$ & 
$\langle a_4/a\rangle$ & [Mg/Fe]$^b$\\
 & & ($h=0.7$) & (km s$^{-1}$) & & $\times 10^3$ & $\times 10^4$ & $\times 10^4$ & $\times 100$ &  \\
\hline
\multicolumn{10}{c}{\bf COMPACT GROUP GALAXIES}\\
HCG 10b & E1 & $-21.40$ & 2.388 & 0.051 &  5.214 &  1.934 &   0.221 &  0.164  & 0.24\\
HCG 14b & E5 & $-20.42$ & 1.983 & 0.025 &  6.002 & $-$1.885 &   0.924 &  0.577  & 0.45\\
HCG 15b & E0 & $-20.25$ & 2.139 & 0.029 &  5.899 &  0.504 &   1.072 & ---  & ---\\
HCG 15c & E0 & $-20.74$ & 2.209 & 0.029 &  5.676 &  1.797 &   0.242 & $-$0.023  & 0.24\\
HCG 19a & E2 & $-22.56$ & 2.262 & 0.033 &  5.897 & $-$0.051 &   0.641 &  0.737  & 0.26\\
HCG 32a & E2 & $-22.32$ & 2.278 & 0.091 &  7.007 & $-$4.092 &   0.153 &  0.429  & 0.25\\
HCG 37a & E7 & $-21.59$ & 2.420 & 0.031 &  6.269 & $-$1.379 &   0.026 & $-$0.535  & 0.21\\ 
HCG 40a & E3 & $-21.48$ & 2.381 & 0.065 &  5.897 &  0.154 &   0.085 &  1.198  & 0.24\\
HCG 44b & E2 & $-19.74$ & 2.263 & 0.025 &  4.785 &  2.181 &   0.817 &  0.274  & 0.16\\
HCG 46c & E1 & $-19.17$ & 1.980 & 0.032 &  6.795 & $-$6.819 &   0.747 &  0.497  & ---\\ 
HCG 51a & E1 & $-21.20$ & 2.250 & 0.017 &  6.064 &  0.057 &  $-$7.783 & $-$0.006  & 0.29\\
HCG 57c & E3 & $-20.86$ & 2.339 & 0.031 &  6.415 & $-$1.068 &   0.255 &  0.823  & 0.17\\ 
HCG 57f & E4 & $-20.98$ & 2.152 & 0.031 &  6.730 & $-$2.685 &   0.337 &  0.538  & 0.20\\
HCG 62a & E3 & $-20.98$ & 2.326 & 0.052 &  5.560 &  0.056 &   0.298 & $-$0.106  & ---\\
HCG 68b & E2 & $-20.39$ & 2.205 & 0.011 &  5.392 & $-$0.129 &  $-$0.533 & $-$0.887  & 0.17\\
HCG 93a & E1 & $-21.87$ & 2.372 & 0.136 &  5.110 &  1.410 &   0.427 & $-$0.289  & 0.24\\
HCG 96b & E2 & $-20.00$ & 2.298 & 0.059 &  5.147 &  1.302 &   1.057 &  0.458  & 0.16\\
HCG 97a & E5 & $-21.09$ & 2.134 & 0.035 &  5.944 & $-$1.478 &  $-$0.736 & $-$0.004  & 0.09\\
\hline
\multicolumn{10}{c}{\bf FIELD GALAXIES}\\
NGC  221 & E2 & $-17.42$ & 1.904  & 0.062 &  4.761 &  1.454 &   1.152 & --- & $-$0.04   \\
NGC  584 & E4 & $-20.63$ & 2.295  & 0.042 &  5.105 &  1.771 &   0.204 &  0.29 & 0.24\\ 
NGC  636 & E3 & $-19.66$ & 2.209  & 0.026 &  5.478 &  0.924 &   0.592 &  0.07 & 0.16\\ 
NGC  821 & E6 & $-20.58$ & 2.282  & 0.110 &  5.056 &  1.920 &  $-$0.113 &  1.16 & 0.26\\
NGC 1700 & E4  & $-21.80$ & 2.378 & 0.043 &  5.694 &  0.664 &   0.154 &  0.48 & 0.15\\
NGC 2300 & SA0 & $-20.87$ & 2.397 & 0.097 &  5.261 &  0.194 &  $-$0.263 & $-$0.39 & 0.39\\
NGC 3377 & E5 & $-19.17$ & 2.132  & 0.034 &  5.546 &  0.691 &   0.584 &  0.81 & 0.22\\
NGC 3379 & E1 & $-20.57$ & 2.304  & 0.024 &  5.043 &  2.032 &  $-$0.008 &  0.08 & 0.26\\
NGC 4552 & E0 & $-20.81$ & 2.386  & 0.041 &  5.092 &  1.343 &   0.092 &  0.02 & ---\\
NGC 4649 & E2 & $-21.54$ & 2.493  & 0.026 &  4.842 &  1.759 &  $-$0.586 & $-$0.41 & ---\\
NGC 4697 & E6 & $-20.97$ & 2.211  & 0.030 &  5.208 &  1.438 &   0.334 &  1.04 & 0.14\\
NGC 7619 & E2 & $-21.86$ & 2.462  & 0.080 &  4.561 &  1.764 &   0.243 &  0.24 & 0.29\\
\hline
\end{tabular}

$^a$~The PCs correspond to the calibrated SEDs in the $3800-4200$\AA\ range.\\
$^b$~Measured within R$_e/2$.
\end{minipage}
\end{table*}

\begin{figure}
\begin{center}
\includegraphics[width=3.5in]{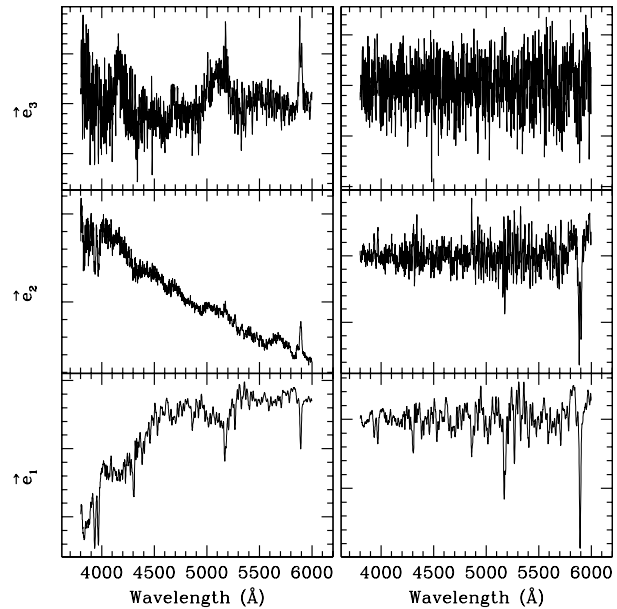}
\end{center}
\caption{First three principal components of the analysis for the flux
calibrated data ({\sl left}) and the continuum subtracted SEDs ({\sl
right}). Notice $\vec{e_1}$ and $\vec{e_2}$ map the average old
populations and a younger stellar component, respectively.}
\label{fig:pcs}
\end{figure}

\section{Principal Component Analysis}

In order to find systematic differences in the stellar populations of
group and field ellipticals we apply principal component analysis
(PCA) to the spectral energy distributions in various ways
(Throughout this paper, the spectra are given as photon flux: F($\lambda$)).
PCA is a
model-independent method which aims at extracting those linear
combinations of the data with the highest variance. Stellar spectra
were the first astrophysical data on which principal component
analysis was applied (Deeming 1964).  PCA has been extensively used in
the analysis of spectra from quasar and galaxy surveys (e.g. Francis
et al. 1992; Connolly et al. 1995; Folkes, Lahav \& Maddox 1996;
Madgwick et al. 2002; Yip et al. 2004). This technique is ideal for
our purposes. Regardless of the responsible physical mechanisms, we
want to find in a robust way systematic differences between the
stellar populations in group and field elliptical
galaxies. Eventually, we compare our results with simple models of
galaxy formation to give physical meaning to these differences.

Our application of PCA is as follows: We have 30 spectra
$\Phi_i(\lambda )$ ($i=\{1\cdots 30\}$) which are defined at a number
of discrete wavelengths $\lambda_j$ ($j=\{1\cdots N\}$). The spectra
are normalized to the same flux integrated along the wavelength region
of interest.  PCA consists of diagonalizing the covariance matrix:
\be
C_{jk}=\frac{1}{30}\Sigma_{i=1}^{30}\Phi_i(\lambda_j)\Phi_i(\lambda_k)\hfil
1\leq j,k\leq N, 
\ee 
so that the eigenvectors ($\{\vec{e}_i\}$) corresponding to the
largest eigenvalues ($\{\nu_i\}$) will carry most of the information
in these spectra.  Notice that other methods use a different
definition for the covariance matrix, most commonly including a
subtraction of the average of the spectra of all galaxies at each
wavelength (see e.g. Ronen, Arag\'on-Salamanca \& Lahav 1999). 
This can be motivated if the observed spectra suffer from strong flux
calibration uncertainties. Our sample, albeit small, has an accurate
flux calibration which allows us to include the information from the
continuum. We have performed tests of the other approach,
namely subtracting the mean from the SEDs, and we find that this
method roughly corresponds to ``promoting'' PC2 to PC1, etc. The correlations
found are unaffected. Henceforth we present results based on PCA
{\sl without} mean subtraction.

One can use the eigenvectors as a basis, so that the components which
correspond to the highest $\nu_i$ will carry most of the
information. It is customary to sort the eigenvectors in decreasing
order of their eigenvalues, and to express these ones as a fractional
variance: $\nu^\prime=\nu_i/\Sigma_j\nu_j$. If the data can be simplified
in a lower dimensional basis, PCA gives a small number of
eigenvectors carrying most of the variance.

\begin{figure*}
\begin{minipage}{15cm}
\begin{center}
\includegraphics[width=5.5in]{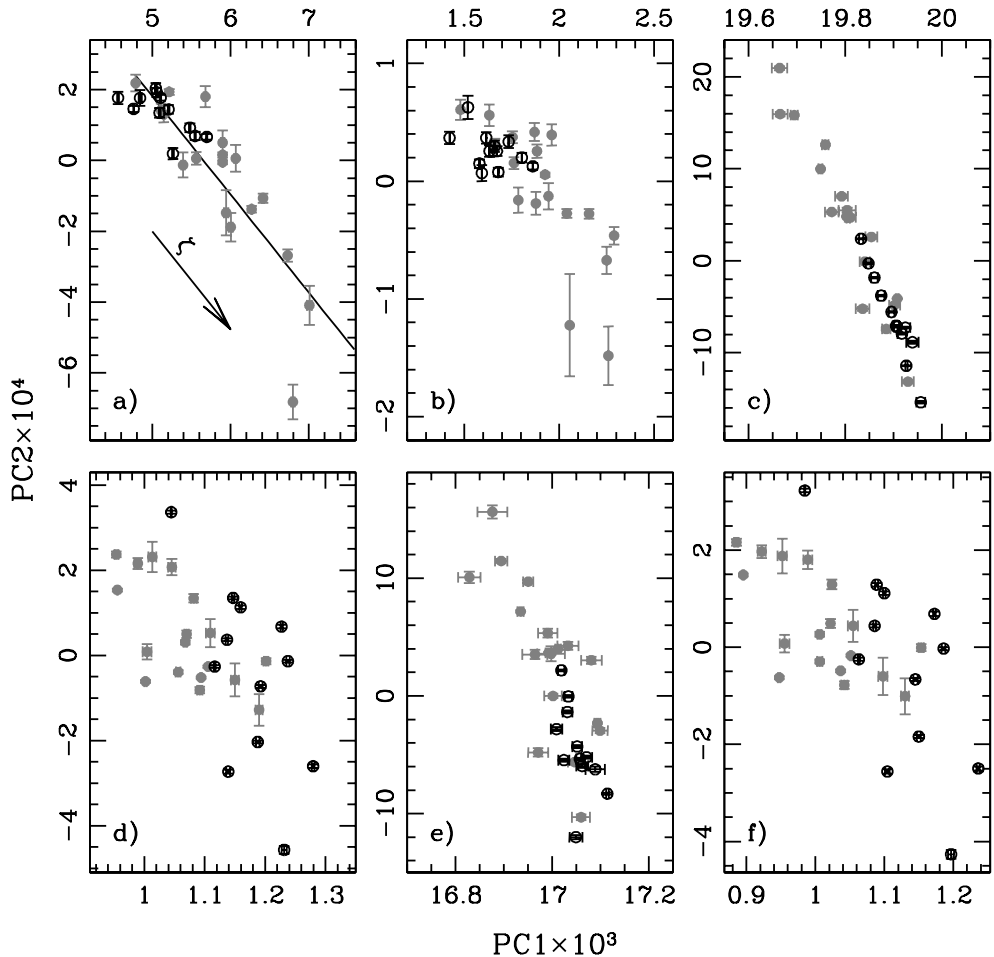}
\end{center}
\caption{Projection of the observed SEDs on the first and second
principal components. HCG and field galaxies are shown as grey dots
and open circles, respectively. Each panel corresponds to: a) Calibrated SEDs
straddling the 4000\AA\ break in the 3800-4200\AA\ range; b) Same as
the previous one, using uncalibrated SEDs; c) Calibrated SEDs over a
wider range of wavelengths: 3800--6000\AA\ ; d) Continuum-subtracted
SED; e) Only use the spectral range used by the Lick indices on the
total SED; f) Same as e) applied to the continuum-subtracted SED. In
all cases there is a clear segregation between group and field
elliptical galaxies. $1\sigma$ error bars shown.}
\label{fig:pc12}
\end{minipage}
\end{figure*}

Figure~\ref{fig:pcs} shows the first three eigenvectors (called
principal components: $\vec{e_1}$, $\vec{e_2}$, $\vec{e_3}$) for two
choices of spectra: The left panels are the principal components
obtained when using the full, flux-calibrated SED. The panels on the
right are those for continuum-subtracted spectra.  The first principal
component corresponds to an old stellar population. As expected,
$\vec{e_1}$ carries most of the information that can be gathered from
an elliptical galaxy, namely a prominent 4000\AA\ break and strong
absorption lines. The second eigenvector, $\vec{e_2}$, gives the next
order of information in our sample, and clearly corresponds to a
younger stellar component. The third component is noisier, but
presents a remarkable feature at $\lambda =5170-5175$\AA\ ,
corresponding to magnesium absorption.  One should not interpret
these SEDs too closely: the emission spike at the wavelength of the
sodium feature ($\lambda =5890-5896$\AA\ ) is just an artifact
caused by the imposed orthogonality of the basis vectors.

Notice that, in contrast to studies applied to surveys of
the entire galaxy population (e.g. Madgwick et al. 2003), our sample
only comprises early-type galaxies. Hence, most of the information
(which will be represented by the first principal component)
corresponds to old stellar populations. Indeed, the normalized
eigenvalues for those $\vec{e_1}$, $\vec{e_2}$, and $\vec{e_3}$ shown
in the left panels of figure~\ref{fig:pcs} are $\nu^\prime_1=0.9943$,
$\nu^\prime_2=0.0013$ and $\nu^\prime_3=0.0007$. This implies that
more than 99\% of the information is carried by the first principal
component, a result consistent with the findings of Eisenstein et
al. (2003) on a sample of $22,000$ luminous early-type galaxies
from SDSS.  This is an ``elegant way'' to show that elliptical
galaxies are dominated by old stellar populations. The aim of this
paper is to explore PCA as a way of extracting small contributions
from younger populations.

In order to robustly determine whether the star formation history of
group and field elliptical galaxies differ in a significant way, we
have to take into account possible systematic effects.  Flux
calibration errors introduce a smooth large-range variation in the
SEDs, possibily mimicking effects of age or metallicity. In order to
explore this effect, we applied PCA to ``variations'' of the spectral
data. Figure~\ref{fig:pcs} shows two such variations, namely the full,
calibrated SED (left) and a continuum subtracted one (right).  We also
considered a reduced spectral range straddling the 4000\AA\ break
(3800--4200\AA ) thus minimising the effect of a flux-calibration
error. Furthermore, we applied PCA to the raw, uncalibrated, unreddened
data in this shorter spectral range. All SEDs were taken with the same
instrument under the same configuration. Hence, we would expect raw
data to carry the safest information about intrinsic differences among
SEDs. The continuum-subtracted SEDs were obtained by subtracting the
boxcar-smoothed spectrum using a 150\AA\ window. Finally, we applied
PCA to two sets of SEDs in which only the spectral regions targeted by
the Lick indices (Trager et al. 1998) were considered. The wavelengths
outside of both central and sidebands were masked out. These indices
were specifically targeted to isolate the spectral regions which are
most sensitive to age or metallicity. After masking out the other regions, 
we expect to reduce noise in the final results. This
technique was applied both to the flux-calibrated SED and to the
continuum-subtracted spectra, as discussed in the next section.

\section{Results}

Figures~\ref{fig:pc12} and \ref{fig:pc23} show the results for the
various methods given above. Once the covariance matrix is
diagonalised, we project the observed SEDs on the first three
principal components. For instance, the first component of the $ith$
galaxy is computed in the following way:

\be 
PC1_i =
\vec{\Phi}_i\cdot\vec{e}_1=\sum_{j=1}^N
\Phi_i(\lambda_j)e_1(\lambda_j).  
\ee 

These projections correspond to the values shown in the figures as
PC1, PC2 and PC3.   Given that PCA diagonalises the covariance
matrix, it is invariant with respect to a sign change in the
projections, e.g.  $\vec{e}_2$ and $-\vec{e}_2$ are equally valid eigenvectors. We
fix this by choosing the sign that always results in an
anticorrelation between PC1 and PC2.  In order to compute the
uncertainty in the principal components, we follow a Monte Carlo
method: For each galaxy we generate 100 spectra by adding noise
compatible with the observed S/N. Each set of ``noisy'' SEDs is presented to PCA
and the projected value for each galaxy from these 100 realisations is
used to determine the error bars, quoted throughout this paper at the
$1\sigma$ level. Notice that the projections of the higher-order
principal components are much smaller, which reflects the amount of
information (in the sense of variance) carried by each of these first
three components.  The remaining higher-order components are noisier
and we do not include them in this analysis. In fact, PCA has filtered
the data so that most of the information from each spectrum can be
reduced to three numbers, namely PC1, PC2 and PC3.  Throughout this
paper galaxies belonging to Hickson Compact Groups are represented as
grey filled dots, whereas the field sample is shown as open circles.

\subsection{Integrated SEDs}

\begin{figure}
\begin{center}
\includegraphics[width=3.5in]{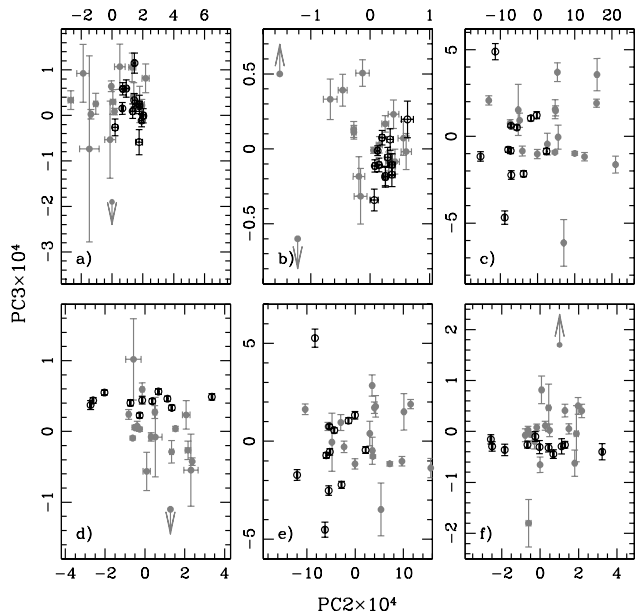}
\end{center}
\caption{Projection of the sample SEDs on the second and third
principal components. Group and field galaxies are shown as
grey dots and star, respectively. The panels correspond to the same
cases as in figure~\ref{fig:pc12}. Arrows are shown for a couple of 
galaxies which appear as outliers: HCG~51a in panels a,b,d,f and HCG~46c
in panel b.}
\label{fig:pc23}
\end{figure}

Figures~\ref{fig:pc12} and \ref{fig:pc23}
correspond to the integrated spectra from each galaxy, i.e. we only
explore galaxy-to-galaxy variations, without any spatial resolution.
Figure~\ref{fig:pc12} shows a remarkable segregation between the group
and field sample for all possible choices of SEDs used: a)
flux-calibrated and de-redenned SED in the 3800--4200\AA\ range; b)
raw data in the same spectral range; c) flux-calibrated and
de-reddened SED in the 3800--6000\AA\ range; d) Continuum-subtracted
SED; e) ``Lick masking'' applied to the flux-calibrated data; f)
``Lick masking'' applied to the continuum-subtracted SEDs.  All
different techniques give a similar segregation in the PC1--PC2 plane.
Field galaxies populate a much smaller range of values of PC1 and PC2
compared to group galaxies. The strong correlation between PC1 and
PC2 gives a hint of the importance of these two components as seen in
figure~\ref{fig:pcs}. PC1 represents old stellar populations, whereas
PC2 tracks younger stars. PCA unfortunately cannot be used to
determine the ``physics'' behind the correlation, but we will see
below that this trend is expected from stellar evolution. The top left panel
of figure~\ref{fig:pc12} also shows the best linear fit to the correlation
found between PC1 and PC2. We can then rotate the coordinate system, so that
one of the axis carries most of the information. This rotated axis defines a
spectral parameter $\zeta$ similar to the $\eta$ parameter in 
Madgwick et al. (2002). However, we must emphasize that the SEDs used
in that case correspond to a wide range of morphologies and the 
mean had to be subtracted in order to minimise uncertainties in 
flux calibration. The connection between $\zeta$ and stellar populations
is discussed in \S5.

Figure~\ref{fig:pc23} shows that PC3 is a rather noisy component (as
well as the higher order ones). Hence, most of the information is
carried by the first and second principal components, and we will
concentrate hereafter on these two. Notice that HCG~51a is shown as an
arrow in panels a, b,d and f, being an outlier with high values of PC3
compared to the rest of the sample. HCG~46c also appears as an outlier in
panel b. Panels d and f -- which correspond to continuum-subtracted
spectra -- show a segregation, with the field sample
featuring a significantly narrower spread in PC3.

\subsection{SEDs in apertures}
\begin{figure}
\begin{center}
\includegraphics[width=3.5in]{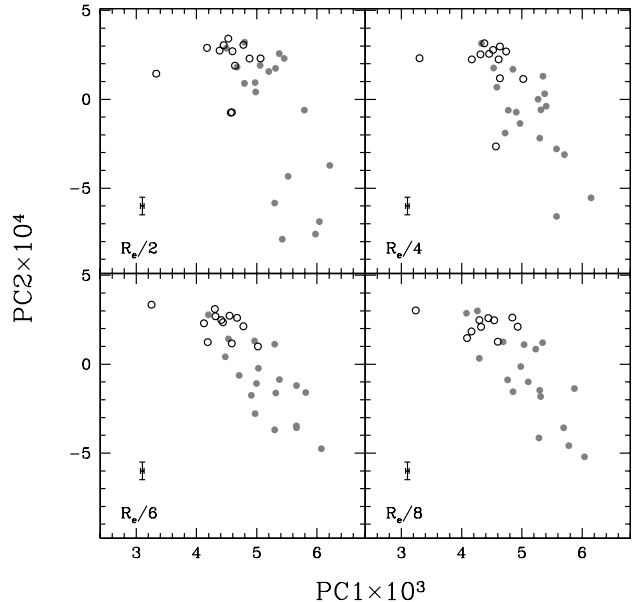}
\end{center}
\caption{The first two principal components are shown taking the
3800-4200\AA\ spectral range for different apertures, chosen with
respect to the half-light radius of each galaxy. A characteristic
error bar is shown}
\label{fig:pcapps}
\end{figure}

The single slit spectra used so far -- obtained over an aperture that
maximises the flux available -- have very high S/N, enabling us to
extract SEDs in smaller apertures.  Figure~\ref{fig:pcapps} presents
the results of PCA applied to four different apertures chosen with
respect to the half-light radius of each galaxy. These PCs correspond
to the analysis of the 3800--4200\AA\ flux-calibrated data, but the
other methods give similar results. The data show a very similar
segregation as in the integrated case, and no significant differences
are found with respect to the choice of aperture. Hence, environmental
differences appear to be related to a global factor. This result is to be
expected if we assume that the formation timescale is short compared
to their stellar ages. The small evolution found in the colour
gradients of elliptical galaxies with redshift suggest these gradients
are mostly a metallicity effect (Tamura \& Ohta 2000; Ferreras et
al. 2005).  Furthermore, integral field unit spectroscopy targeting
age-sensitive spectral features such as Balmer absorption show a
rather homogeneous structure in kinematically distinct components
(Davies et al. 2001, however see McDermid et al.  2005). Our results
reveal that the differences between group and field early-type
galaxies are global and do not relate to significant differences in
their gradients.

\subsection{Structural parameters}

Structural differences should be expected between group and field
galaxies.  The higher density but low velocity dispersions found in
compact groups favour an increased interaction or merger rate
among galaxies. Hence, elliptical galaxies in the cores of HCGs should
reveal dynamical signatures from recent encounters (Proctor et
al. 2004; Coziol et al. 2004)

We measured their isophotal shapes with the IRAF ELLIPSE routine
(Jedrzejewski 1987) on the survey images acquired by Hickson (1994) in
the $R$ band with the 3.6m Canada-France-Hawaii Telescope
(CFHT). These images are publicly available on the NASA/IPAC
Extragalactic Database (NED).  ELLIPSE allows the isophote center,
ellipticity and position angle to vary at each iteration.  Once a
satisfactory fit is found, it computes the sin and cos 3$\theta$ and
4$\theta$ terms, which describe the deviations of the isophotes from
pure ellipses by means of the following Fourier expansion:

\be
\begin{array}{cc}
\Delta r(\theta)/r(\theta) = & 
(a_3/a)\cos (3\theta) + (a_4/a)\cos (4\theta) +\\
 & \\
 & + (b_3/b)\sin (3\theta) + (b_4/b)\sin (4\theta),
\end{array}
\ee

\noindent
where $\theta$ is the position angle, $r$ the distance from the galaxy
centre and $a$ and $b$ the semi-major and semi-minor axes.  The
$a_4/a$ and $a_3/a$ parameters, where $a$ is the semi-major axis of
the isophote, measure the deviations of the isophote from a pure
ellipse.  In particular, $a_4/a$ measures the deviations which are
symmetric with respect to the galaxy center (i.e. found along the
isophote every $90^o$).  Positive (negative) values are obtained for
disky (boxy) galaxies. Simulations show that boxy galaxies are
remnants of massive mergers, whereas disky ellipticals are related to
a larger mass difference between the merged progenitors (cf. Khochfar
\& Burkert 2005, Jesseit et al. 2005). 

We ran ELLIPSE on the images taken in the $R$ filter, leaving as free
parameters the position of the centre, the ellipticity (defined as 
$1-b/a$, where $a$ and $b$ are the projected major and minor axes) and
the position angle of the major axis and assuming a logarithmic step
along the semi-major axis. We computed the average $\langle
a_3/a\rangle$ and $\langle a_4/a\rangle$ parameters of each galaxy
following the prescriptions of Bender et al. (1988) and in the R
filter. The method in question consists of averaging the $a_3/a$ and
$a_4/a$ values measured at R$_i \leq r \leq $R$_o$ weighted by their
errors and by the counts in the isophote at each isophotal mean radius
$r$. Here, R$_o$ is 1.5 times the half-light radius (R$_{\rm e}$)
derived from the surface brightness radial profile and R$_i$ is the
radius corresponding to 3 times the FWHM of the PSF in the $R$ filter
-- FWHM(PSF)$\simeq 4$ pixels hence $1.6^{\prime\prime}$.  
The resulting mean values\footnote{The $\langle a_3/a\rangle$ values
were also computed. However, this parameter was found to be --
within error bars -- consistent with zero for all galaxies. 
$\langle a_3/a\rangle$ measures
the non-symmetrical deviations occurring at every $120^o$, possibly
due the presence of dust lanes and clumps.}
$\langle a_4/a\rangle$ are reported in
table~\ref{tab:sample}, together with the mean isophotal parameters of
the field galaxies taken from Bender et al. (1988, 1989).

Figure~\ref{fig:pca4} compares PC1, PC2 and PC3 with the structural
parameter $a_4$ described above. The number of galaxies is too small to
draw statistically significant conclusions. Most of the galaxies have
either disky or very close to elliptical isophotes, although the ones
with the highest values of PC1 and PC2 -- which are mostly group
galaxies -- feature disky isophotes. PC3 presents a more significant
correlation, an issue which is explored below in connection with
velocity dispersion.

\subsection{[Mg/Fe] abundance ratios and velocity dispersion}

Abundance ratios such as [Mg/Fe] are very sensitive tracers of the
duration of star formation. Magnesium is an alpha element, mainly
produced in core-collapse supernovae, whereas most of the iron is
synthesized during the explosion of a type Ia supernova (see e.g.
Iwamoto et al. 1999). The latter
are triggered by the accretion of gas on to a CO white dwarf from a
close companion. Hence, the time for the onset of type Ia is delayed
with respect to core collapse supernovae by as much as 0.5--1 Gyr (see
e.g. Matteucci \& Recchi 2001). The enhanced [Mg/Fe] ratios commonly
found in massive elliptical galaxies (e.g. Trager et al. 2000) suggest
that the bulk of the stellar component was formed roughly over a
dynamical timescale.  The scaling relation between velocity
dispersion and [Mg/Fe] further suggests that lower mass ellipticals
have a lower efficiency of star formation (Ferreras \& Silk 2003;
Pipino \& Matteucci 2004, however see Proctor et al. 2004).

\begin{figure}
\begin{center}
\includegraphics[width=3.5in]{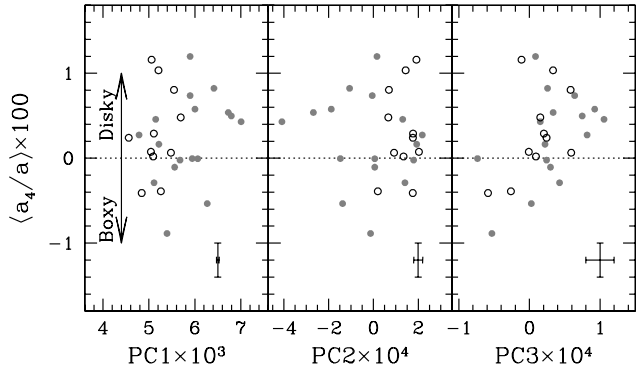}
\end{center}
\caption{Comparison between the first three principal components
and the structural parameter $a_4$  (see text for details).
Characteristic error bars are shown.}
\label{fig:pca4}
\end{figure}

\begin{figure}
\begin{center}
\includegraphics[width=3.5in]{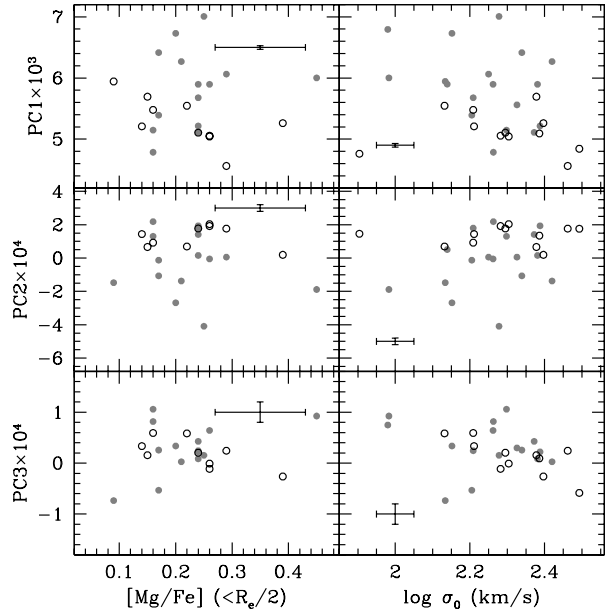}
\end{center}
\caption{The first three principal components are compared with respect
to the [Mg/Fe] abundance ratios measured within R$_e/2$ (left) and the 
central velocity dispersion (right).}
\label{fig:pcmgfe}
\end{figure}

Figure~\ref{fig:pcmgfe} compares the PCA results with both the [Mg/Fe]
ratios within R$_e/2$ (left) and the central velocity dispersion
($\sigma_0$; right). The [Mg/Fe] values -- also presented in
table~\ref{tab:sample} -- were obtained using the Thomas, Maraston \&
Bender (2003) stellar population models, which are sensitive to the
abundance ratios. The data do not show any strong correlation, in
remarkable contrast to the one found in figure~\ref{fig:pc12}. 
However, notice PC3 ({\sl bottom panel}) shows a correlation
with respect to velocity dispersion. The third component presents a
feature around the magnesium absorption region $\lambda
=5170-5175$\AA\ . The trend found in this figure can thereby be
related to the Mg-$\sigma$ relation (see e.g. Bernardi et
al. 1998). This trend is similar to the one found in
figure~\ref{fig:pca4} between $\langle a_4/a\rangle$ and PC3, although
these two correlations are ``related'': galaxies with lower velocity
dispersions are known to have diskier isophotes (Bender et
al. 1989). We could expect a similar trend with the abundance ratios,
as these are also correlated with velocity dispersion (see e.g. Trager
2000).  Unfortunately, the error bars in [Mg/Fe] are too large, and no
``follow-up'' of this trend can be seen.

\section{Giving physical meaning to PCA}

\begin{figure}
\begin{center}
\includegraphics[width=3.5in]{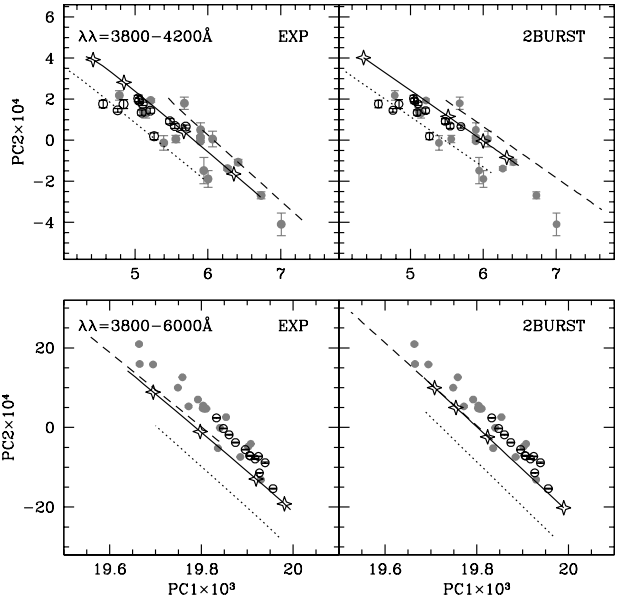}
\end{center}
\caption{PCA and star formation histories (I): The top and bottom
panels show PC1 and PC2 for the calibrated SEDs in the short
(3800--4200\AA\ ; top) and long (3800--6000\AA\ ; bottom) wavelength
range. The error bars are given for 100 realizations of the SED with
the noise characteristics taken from the observations.  The lines are
projections of synthetic models at fixed metallicity, formed at
$z_F=3$. The left panels (EXP model) span a range of exponentially
decaying star formation timescales between $\tau_{\rm SF}=0.1$ and
$5$~Gyr, whereas the panels on the right (2BURST model) fix 
$\tau_{\rm SF}=0.5$~Gyr and superimpose a young population of age 
1~Gyr with a mass fraction $f_Y$ between $0$ and $0.2$.  The solid and
dashed lines correspond to dustless models with metallicity $2Z_\odot$
and $Z_\odot /2$, respectively. The dotted lines show the prediction
for solar metallicity and E(B$-$V)$=0.2$~mag.}
\label{fig:pcsfh1}
\end{figure}

In order to relate this environmental segregation to actual
differences in star formation history we need to project the
principal components on to spectra corresponding to stellar
populations whose ages and metallicities are known (Ronen et al. 1999;
Madgwick et al. 2003). We generate two sets of models
which describe the star formation history in terms of a single free
parameter. These models give a distribution of stellar ages that is
used to convolve simple stellar populations from the synthesis models
of Bruzual \& Charlot (2003). A Chabrier initial mass function (IMF)
was used, although the result does not change significantly with
respect to the choice of IMF. The composite SEDs are subsequently
degraded to the instrumental resolution and velocity dispersion of a
typical galaxy from our sample. The spectra are finally projected on
to the basis vectors obtained by the application of PCA on the {\sl
real} data, and the results are shown in figures~\ref{fig:pcsfh1} and
\ref{fig:pcsfh2}. The two sets of models assume a fixed metallicity
and dust content, and describe the star formation rate in the
following way:
\begin{enumerate}
\item EXP: Exponentially decaying star formation rate: 
SFR$(t)\propto\exp [-(t-t(z_F))/\tau_{\rm SF}]$. The process starts 
at a redshift $z_F=3$. The free parameter is the timescale
($\tau_{\rm SF}$), assumed to be in the range $0.1-5$~Gyr.
\item 2BURST: An old population is generated using an
exponentially decaying star formation rate starting at $z_F=3$ with a 
timescale $\tau_{\rm SF}=0.5$~Gyr following the notation in (i). 
A second population is added with an age of 1~Gyr and the
same metallicity as the old population. The free parameter is $f_Y$, which
represents the stellar mass fraction in young stars.
\end{enumerate}

Figure~\ref{fig:pcsfh1} shows the (PC1,PC2) diagram obtained with the calibrated SEDs over a
$\pm 200$\AA\ spectral range that straddles the 4000\AA\ break (top) and
over a wider range of wavelengths ($3800-6000$\AA\ ; bottom). The
error bars are obtained by a Monte Carlo sampling of the SEDs. For
each of the 100 realizations noise was added to each pixel according
to the observed S/N. The error bars give the $1\sigma$ confidence
levels.  The lines are the projections of the models described
above. The solid and dashed lines are dustless models with metallicity
$2Z_\odot$ and $Z_\odot /2$, respectively, whereas the dotted line
corresponds to solar metallicity and a dust reddening of E(B$-$V)$=0.2$
using the $R_v=3.1$ extinction model of Fitzpatrick (1999). The lines
in the EXP models span a range of timescales $\tau_{\rm
SF}=0.1-5$~Gyr. The stars (from the bottom left corner) correspond to
$\tau_{\rm SF}=1, 2, 3$ and $4$~Gyr.  The lines in the 2BURST models
span a range of young stellar mass fractions $f_Y=0-0.2$, with the
stars (from bottom-left corner) at $f_Y=\{ 0,0.05,0.01,0.15\}$.

Figure~\ref{fig:pcsfh2} shows the model prediction as a function of
the free parameter, namely the star formation timescale in the EXP
model (left) and the mass fraction in young stars for the 2BURST case
(right). The same notation as in figure~\ref{fig:pcsfh1} applies to
the solid, dotted and dashed lines. Instead of showing
components PC1 and PC2 independently, we use a linear combination
which captures the correlation between them. We rotate the (PC1,PC2)
coordinate system so that one of the rotated axis follows the least
squares fit to the data from panel a) in figure~\ref{fig:pc12}
(calibrated SEDs over $\Delta\lambda =3800-4200$\AA\ ). Hence, we
define:
\be
\zeta\equiv 0.36 {\rm PC1} - {\rm PC2}.
\ee
This method is analogous to the definition of the spectral parameter
$\eta$ in Madgwick et al. (2002). However, in our
paper the range of spectra is reduced to early-type galaxies, and we
use information from the stellar continuum. Hence, we cannot
establish a one-to-one correspondence between $\zeta$ and $\eta$.

For comparison, the histogram of the field and compact group samples
are shown as a black and grey line, respectively.  The model
predictions are given as a function of $\zeta$. Our HCG sample
features a wider distribution, including an important number of
galaxies with values of $\zeta$ that corresponds to the presence
of younger stars, either as a more extended SFH (in the EXP case) or
as a higher value of $f_Y$ (2BURST model).

\section{Discussion}

In order to find environmental effects, compact groups are arguably
the best places to look for recent phases of formation.  Their
moderate velocity dispersion and relatively high densities result in a
fertile scenario for galaxy mergers. Recent work targeting the
Fundamental Plane (de~la~Rosa, et al. 2001a) has not found significant
differences with respect to early-type galaxies in the field.  
On the other hand, Proctor et al. (2004) used spectral
indices to suggest that the ages of HCG galaxies were more similar to
cluster galaxies.  Environmental effects were not found between field
and cluster galaxies by way of the Mg--$\sigma$ relation (Bernardi et
al. 1998) or other kinematic and chemical signatures (Ziegler et
al. 2005). The infamous age-metallicity degeneracy and the
uncertainties underlying the translation from observed line strengths
to actual abundance ratios prevent us from an accurate assessment of
the difference in the star formation of early-type systems.

\begin{figure}
\begin{center}
\includegraphics[width=3.5in]{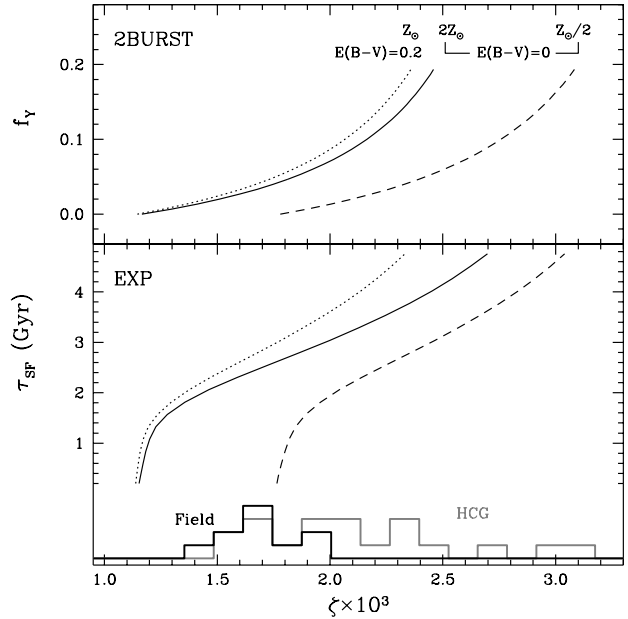}
\end{center}
\caption{
PCA and star formation histories (II): The histograms 
are the data measurements of group (grey) and field elliptical
galaxies (black) according to parameter $\zeta =0.37PC1-PC2$ 
(see text for details). 
The curved lines are predictions according to the same two
models described in figure~\ref{fig:pcsfh1}. Notice that galaxies in
Hickson groups have a wider range of $\zeta$, suggesting 
a larger range of stellar ages.}
\label{fig:pcsfh2}
\end{figure}

The data used in this paper comprises a consistent sample of
elliptical galaxies in HCGs and in the field. The same telescope,
instrument and configuration has been used in our galaxies. Hence, a
model-independent approach such as principal component analysis is
justified. The advantage of PCA over other techniques is that we
maximise the information (in the sense of variance) extracted from the
data without any reference to models. In this way we have found that
no matter how we present the data to PCA -- over a small range of
wavelengths; without calibration; with continuum subtraction;
selecting only the Lick/IDS spectral regions -- we always find the
same segregation between field and group samples. {\sl HCG elliptical
galaxies span a wider range of principal components}. This result, on
its own, proves that early-type galaxies in compact groups have
different star formation histories. Interestingly enough, more
``traditional'' observables such as the Fundamental Plane, structural
parameters tracking the departure from elliptical isophotes, or
[Mg/Fe] abundance ratios do not seem to be sensitive enough to these
changes (see e.g. de la Rosa et al. (2001ab), Proctor et
al. (2004) and Zepf \& Withmore (1993). The segregation in (PC1,PC2)
space is also found for the SEDs measured in different apertures, from
$R_e/8$ out to $R_e$, implying that this effect does not have to do
with radial gradients, but with a more global difference in the star
formation histories.
 
Figures~\ref{fig:pca4} and \ref{fig:pcmgfe} explore the 
correlation of the PCs with other observables such as the 
structural parameter $a_4$ --  which gives the diskyness/boxyness of the isophotes,
velocity dispersion, or [Mg/Fe]. Remarkably, we find no correlation with
either PC1 or PC2, but -- within the limitations of our small sample --
PC3 appears to be correlated with these ``dynamically-related'' 
observables.

The weakness of PCA is that it does not suggest any ``physics'' to
explain this segregation. Figure~\ref{fig:pc12} cleary shows that our
field galaxies have similar star formation history, whereas HCG
galaxies present a more complex distribution. In order to interpret
these results, simple models spanning a range of ages and
metallicities were generated and projected on the basis vectors given
by the {\sl real} data. Both models (an exponentially decaying star
formation rate or a two-burst model) suggest younger ages for a
fraction of the sample in HCGs. In contrast, the field sample is
consistent with old stellar populations. Figure~\ref{fig:pcsfh2}
implies that the wider distribution of the spectral parameter $\zeta$
in group galaxies is caused by a wider range of stellar ages.  
With respect to the correlation of PC3 with velocity dispersion or
$a_4$, this component (see figure~\ref{fig:pcs}) has a feature around
the magnesium absorption region. Hence, the correlations found 
tell us that higher velocity dispersions correspond to galaxies with 
metal-rich populations (i.e. the Mg--$\sigma$ relation, Bernardi et al. 1998)
with boxy isophotes (Bender et al. 1989).

More work and larger samples are needed to exploit the full power of
PCA, but the results we present here should illustrate both the
complexity of extracting star formation histories from unresolved
stellar populations and the potential of ``non-traditional'' statistical
techniques towards the understanding of galaxy formation.

\section*{Acknowledgments}
The anonymous referee is gratefully acknowledged for very
useful comments.
We would like to thank Vivienne Wild and Paul Hewett for useful comments.
This research has made use of the NASA/IPAC
Extragalactic Database (NED) which is operated by the Jet Propulsion
Laboratory, California Institute of Technology, under contract with
the National Aeronautics and Space Administration.


\end{document}